\documentstyle[12pt]{report} \voffset=-30mm
\begin{document}
\setlength{\baselineskip}{22pt}
\begin{center}
{\Large \bf Fractals, Multifractals and the Science of complexity} 
 
M. K. Hassan
 
Department of Physics
 
Brunel University
 
Uxbridge, Middlesex
 
UB8  3PH
 
United Kingdom
\end{center}

\vspace{5mm}

Describing natural objects by geometry is as old as science 
itself,   traditionally this has involved the use of Euclidean lines, 
rectangles,  cuboids,  spheres, and so on. But, nature is not restricted 
to Euclidean  shapes. More than twenty years  ago Benoit B. Mandelbrot 
observed that  ``Clouds are not spheres, mountains are not cones, 
coastlines are not  circles, bark is not smooth, nor does lightning 
travel in a straight  line". Most of the  natural objects we see around 
us are so complex in  shape as to deserve being called geometrically 
chaotic. They appear impossible to  describe mathematically and used to 
be regarded as the `monsters of mathematics'. 

In 1975, Mandelbrot introduced the concept of fractal geometry to 
characterize these monsters quantitatively and to help us to appreciate 
their underlying regularity [1]. Fractals are more than  brightly-coloured, 
computer generated patterns. The  coastline of an island, a river 
network, the structure of a cabbage or broccoli, or the network of 
nerves and blood vessels in the normal human retina can be best described 
as fractals. Yet, more than twenty years after they were first introduced 
there is no generally-accepted definition of a fractal, although it can 
be defined loosely  as a shape made of parts similar to the 
whole in some sense. 

The simplest way to construct a fractal is to repeat a given operation 
over and over again deterministically. The classical Cantor set is a 
simple text book example of such a fractal. It is created by dividing a 
line into $n$ equal pieces and removing $(n-m)$ of the parts created and 
repeating the process with $m$ remaining pieces {\it ad infinitum} [1,2]. 
However, fractals that occur in nature occur through continuous kinetic 
or random processes.  Having realized this simple law of nature, we can 
imagine selecting  a line randomly at a given rate, and dividing it 
randomly, for example. We can further tune the model to determine 
how random is random. Starting with an infinitely long line we obtain   an 
infinite number of points whose separations are determined by the initial 
line and the degree of randomness with which intervals were selected. The 
properties of these points appear  to be statistically self-similar and 
characterized by the fractal  dimension, which is found to increase with 
the degree of  increasing order and reaches  it's maximum value in the 
perfectly ordered pattern. 

Recently, this idea has been extended  to two dimensions to 
understand fractals in nature that have both size and shape. That is, we 
divide a rectangle randomly into four smaller rectangles and randomly remove 
one of the pieces.  We continue the process {\it ad infinitum} 
with the remaining pieces. In this case it seems that we cannot 
describe the phenomenon  by a single fractal dimension - infinitely 
many are required [3]. Such 
phenomena are called {\it multifractal} and have become a very active 
research  area spanning many disciplines. Physically, it 
means that it is possible to partition the resulting system into subsets 
such  that each subset is a fractal with its own characteristic dimension. 
In this process a new feature appears: that the support on 
which different subsets can be distributed is itself a fractal with one of 
infinitely many possible dimensions. That is, a single 
experiment for a longer time will not give any averaged 
quantity of interest with a good accuracy, but a large number of 
independent experiments are required. Typically, multifractal 
patterns appear in  systems that develop in far from equilibrium and 
that do  not yield a minimum energy configuration, such as diffusion limited 
aggregation, or a metal foil grown by electro-deposition. 

Most of our knowledge about fractals comes from computer simulations, but 
creating fractals  using models of fragmentation process are simple and  
analytically tractable.  These models can describe  
the patterns that arise in random sequential deposition of  a  mixture of 
particles  with a continuous distribution of sizes on a finite substrate. 
In the case of the deposition of particles of a definite size  the system 
clearly reaches a  jamming limit when it is impossible to place further 
objects without overlapping because of its strong non-Markovian and 
non-ergodic nature. With  a continuous distribution of sizes the system 
does not reach a jamming 
limit, but instead creates a scale invariant pattern that can be 
described as a fractal [4,5] since the system gains its ergodic nature.
 
Is it possible to say  when we might  expect a system to exhibit random 
fractal behaviour?  So far, there does not seem to be a  unique answer 
to the question. It appears that,  whenever we  hopelessly fail to produce 
an identical copy of a system under the same initial 
condition, but each copy has the same generic form, we find a fractal object.
Two snowflakes never appear  the same, but due to their generic form even 
a child can recognize them.  The conclusion may be that  creating a 
complex shape is much simpler than it appears at first sight!

\vspace{5mm}

{\large \bf References}

\vspace{5mm}

\noindent
[1] Mandelbrot B B, {\it The Fractal Geometry of Nature} (Freeman, 
San Francisco 1982)

\noindent
[2]  Hassan M K and Rodgers G J  Physics Letters A  {\bf 208} 95
         
[3]   Hassan M K and Rodgers G J (to appear in Physics letters 
A, 1996)              

[4] Brilliantov N V Andrienko Y A Krapivsky P L and Kurths J 1996 Phys. 
Rev. Lett. {\bf 76} 4058 

[5] Hassan M K `Comment on Ref. [4]' submitted to Phys. Rev. Lett.

\end{document}